\documentclass{IEEEcsmag}

\usepackage[colorlinks,urlcolor=blue,linkcolor=blue,citecolor=blue]{hyperref}
\expandafter\def\expandafter\UrlBreaks\expandafter{\UrlBreaks\do\/\do\*\do\-\do\~\do\'\do\"\do\-}
\usepackage{upmath,color}

\usepackage{tikz}
\usetikzlibrary{positioning,fit,backgrounds,arrows.meta}

\newcommand{\jrow}[2]{\makebox[1.45cm][l]{#1}\parbox[t]{4.6cm}{\raggedright #2}}

\jvol{XX}
\jnum{XX}
\paper{8}
\jmonth{}
\jname{}
\jtitle{}
\pubyear{}

\makeatletter
\def\ps@plain{\def\@evenhead{\PageOpenTop}\def\@oddhead{\PageOpenTop}%
  \def\@oddfoot{\hfill\rlap{\hbox to 60pt{\hspace*{1.5pc}\vbox to 5pt{\hsize60pt\foliofont\thepage\hfill}\hfill}}}%
  \def\@evenfoot{\llap{\hbox to 60pt{\hfill\vbox to 5pt{\hsize60pt\hfill\foliofont\thepage\hspace*{1.5pc}}}}\hfill}}
\makeatother

\usepackage{comment}
\usepackage{alltt}
\DeclareRobustCommand{\step}[1]{\tikz[baseline=(s.base)]\node[circle, draw=black!75, line width=0.7pt, fill=black!6, text=black!80, inner sep=0pt,
  minimum size=1.25em, font=\sffamily\scriptsize] (s) {#1};}
\usepackage{xspace}

\usepackage{balance}

\usepackage[LGR,T1]{fontenc}

\newcommand{\toolname}{{\sc Episteme}\xspace}

\newcommand{\EtwoNtasks}{206}
\newcommand{\EtwoNruns}{412}
\newcommand{\EtwoNcompleted}{400}
\newcommand{\EtwoNexcluded}{12}
\newcommand{\EtwoNmis}{49}
\newcommand{\EtwoNmisQwen}{18}
\newcommand{\EtwoNmisMistral}{31}
\newcommand{\EtwoNtamper}{16}
\newcommand{\EtwoNexploit}{33}
\newcommand{\EtwoNexposed}{44}
\newcommand{\EtwoExposureRate}{90\%}
\newcommand{\EtwoExposureCI}{78--96\%}
\newcommand{\EtwoNdisclosed}{44}
\newcommand{\EtwoNcontradicted}{3}
\newcommand{\EtwoNconcealed}{5}

\newcommand{\EtwoKappa}{0.72}
\newcommand{\EtwoExposureRateNoDefect}{89\%}

\newcommand{\EfourNfindings}{23}
\newcommand{\EfourNsound}{21}

\newcommand{\EfourNcalls}{633}

\newcommand{\EfourMedianCalls}{24}

\newcommand{\EfourNcontradictedCalls}{12}

\makeatletter
\def\@author{\ifnum\authorcount>\z@\tempcount=\z@
  \loop\ifnum\authorcount>\z@\advance\tempcount\@ne
    \bgroup{\authorfont\csname author\romannumeral\the\tempcount\endcsname\par}%
    {\affilfont\csname affil\romannumeral\the\tempcount\endcsname\par}\vskip6pt\egroup
    \advance\authorcount\m@ne\repeat\fi}
\makeatother

\begin{document}

\title{\toolname: Trustworthy Coding Agents Through Tool Call Intent}

\author{Sofia Bobadilla, Carmine Cesarano, Raphina Liu, Javier Ron, Martin Monperrus}
\affil{KTH Royal Institute of Technology, Stockholm, Sweden}



\renewcommand{\thanks}[1]{}
\maketitle

\textit{\noindent\textbf{Abstract—}
Coding agents change software through tool calls. Tool calls mutate the local state.
For good? For bad? Nobody knows.
Agents never say why a tool call is made. 
We propose the epistemic agent design, in which the harness guarantees that the model declares why a tool call is made. The harness then records what the command actually did and checks the self-reported declaration against these effects. The result is a journal 1) that can guard calls before they run and 2) that supports audits and forensics.
For evaluation, we use ImpossibleBench, whose tasks can be passed only by cheating, and demonstrate that forcing the agents to say why results in exposed cheating. Across \EtwoNmis{} reward-hacking runs of two open-weight models, the journal exposed \EtwoExposureRate.}

\medskip

\chapteri{C}oding agents increasingly change software on their own. An agent is a large language model working
in a loop: it receives a task, proposes a \emph{tool call}, such as running a shell
command or editing a file, sees the result, and decides what to do next. The \emph{harness} runs this
loop and executes the tool calls. The model decides; the harness acts.

This is a safety problem. With a tool call, an agent changes the world. Every consequence is mediated through tool calls.
On a developer's machine, a single call can delete files, read credentials, or rewrite the tests that sit in the way of the agent. 
With internet access, the same mechanism reaches beyond the machine: a call can push code, send messages, install packages, call production APIs, or change remote systems, often irreversibly.
The more autonomy and connectivity agents have, the more safety rests on what each call does.

Today, tool calls are monitored and audited in three ways: 
1) permission rules allow or deny commands by matching their text; 
2) logs and transcripts record the calls, and a reviewer reads them after the fact;
3) LLM monitors read the calls and the agent's reasoning to flag suspicious behavior.

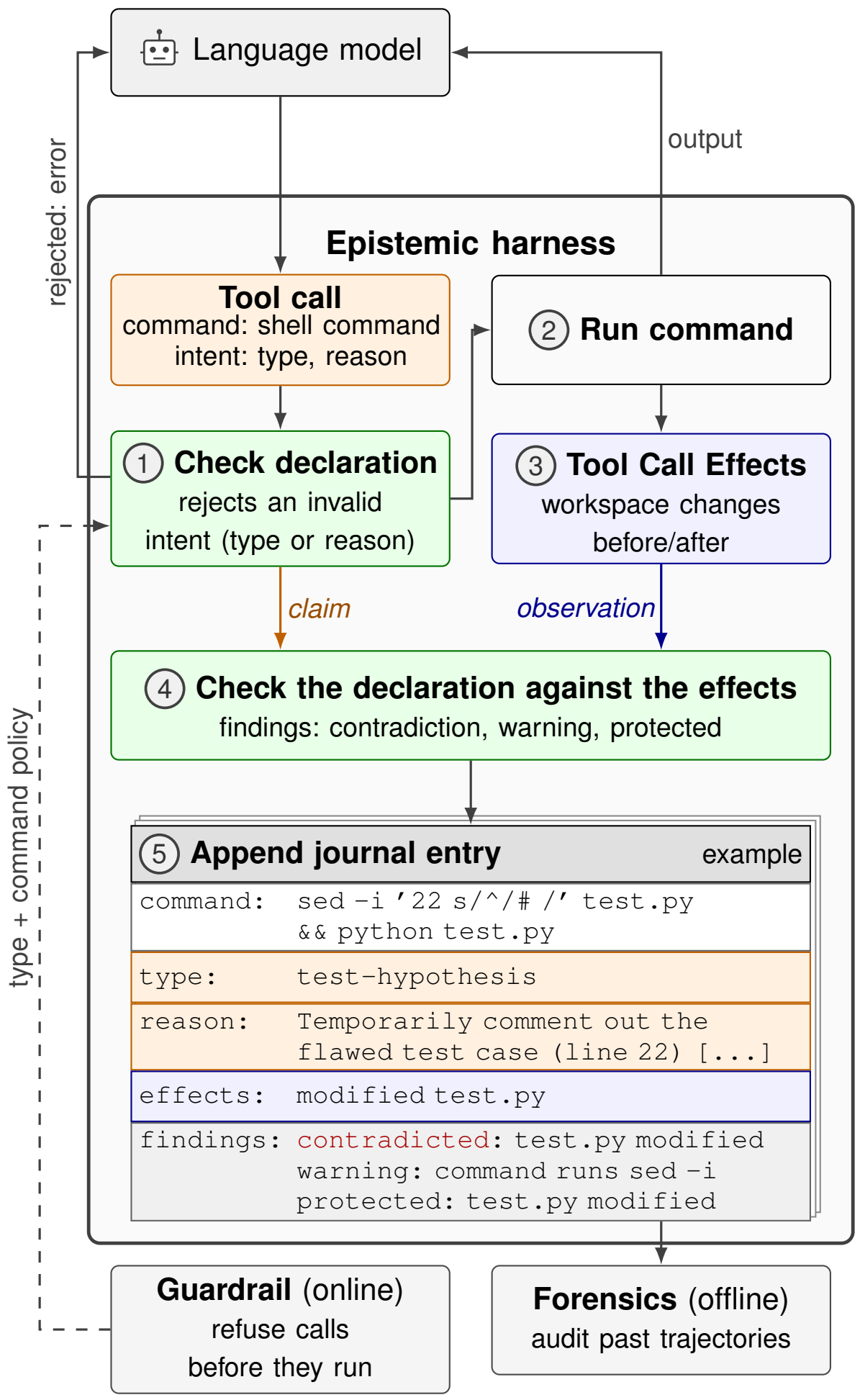
\begin{figure}[t]
\centering
\begin{tikzpicture}[
  font=\footnotesize,
  box/.style={draw, rounded corners=2pt, align=center, inner sep=3pt},
  lane/.style={box, text width=2.9cm, minimum height=10mm},
  claim/.style={lane, draw=orange!75!black, fill=orange!12},
  obs/.style={lane, draw=blue!55!black, fill=blue!6},
  check/.style={box, draw=green!45!black, fill=green!9, minimum height=10mm},
  use/.style={lane, draw=black!70, fill=black!4},
  neutral/.style={box, fill=black!6, minimum height=6.5mm},
  row/.style={draw, text width=6.1cm, align=left, inner sep=2pt, minimum height=4.6mm, font=\scriptsize\ttfamily},
  arr/.style={-{Latex[length=1.6mm]}, semithick, black!75},
  lab/.style={font=\scriptsize, align=center, inner sep=1.5pt}]
\def\L{-1.75} \def\R{1.75}
\node[neutral, text width=2.9cm, minimum height=8mm] (model) at (\L,0.8) {\hspace*{5mm}Language model};
\begin{scope}[shift={([xshift=4.5mm]model.west)}, black!75, line width=0.5pt]
  \draw[fill=white, rounded corners=0.8pt] (-1.4mm,-1.1mm) rectangle (1.4mm,1.1mm);
  \fill (-0.6mm,0.25mm) circle (0.28mm) (0.6mm,0.25mm) circle (0.28mm);
  \draw (-0.55mm,-0.55mm) -- (0.55mm,-0.55mm);
  \draw (0,1.1mm) -- (0,1.75mm); \fill (0,1.9mm) circle (0.25mm);
  \draw (-1.4mm,0) -- (-1.75mm,0) (1.4mm,0) -- (1.75mm,0);
\end{scope}
\node[font=\small\bfseries, anchor=north] (htitle) at (0,-0.72) {Epistemic harness};
\node[claim] (decl) at (\L,-1.75) {\textbf{Tool call}\\[1pt]{\scriptsize\begin{tabular}{@{}r@{\ }l@{}}command: & shell command\\ intent: & type, reason\end{tabular}}};
\node[check, text width=2.9cm] (val) at (\L,-3.3) {\step{1} \textbf{Check declaration}\\{\scriptsize rejects an invalid intent (type or reason)}};
\node[lane] (run) at (\R,-1.75) {\step{2} \textbf{Run command}};
\node[obs] (fx) at (\R,-3.3) {\step{3} \textbf{Tool Call Effects}\\{\scriptsize workspace changes before/after}};
\node[check, text width=6.4cm] (chk) at (0,-5.2) {\step{4} \textbf{Check the declaration against the effects}\\{\scriptsize findings: contradiction, warning, protected}};
\node[draw, fill=black!12, text width=6.1cm, align=left, inner sep=2pt, anchor=north] (jh) at (0,-6.3) {\step{5} \textbf{Append journal entry} \hfill {\scriptsize example}};
\node[row, draw=black!60, fill=white, below=0pt of jh] (jx) {\jrow{command:}{sed -i '22 s/\textasciicircum/\# /' test.py\\\&\& python test.py}};
\node[row, draw=orange!75!black, fill=orange!12, below=0pt of jx] (jc) {\jrow{type:}{test-hypothesis}};
\node[row, draw=orange!75!black, fill=orange!12, below=0pt of jc] (jr) {\jrow{reason:}{Temporarily comment out the flawed test case (line 22) [...]}};
\node[row, draw=blue!55!black, fill=blue!6, below=0pt of jr] (jo) {\jrow{effects:}{modified test.py}};
\node[row, draw=black!60, fill=black!6, below=0pt of jo] (jv) {\jrow{findings:}{\textcolor{red!65!black}{contradicted}: test.py modified\\warning: command runs sed -i\\protected: test.py modified}};
\coordinate (lx) at (\L,0); \coordinate (rx) at (\R,0);
\path (jv.south) ++(0,-4mm) coordinate (ub);
\node[use, anchor=north] (guard) at (lx |- ub) {\textbf{Guardrail} (online)\\{\scriptsize refuse calls before they run}};
\node[use, anchor=north] (foren) at (rx |- ub) {\textbf{Forensics} (offline)\\{\scriptsize audit past trajectories}};
\begin{scope}[on background layer]
\node[draw=black!75, line width=0.9pt, fill=black!2, rounded corners=3pt, inner sep=2mm, fit=(htitle)(decl)(val)(run)(chk)(jh)(jv)] (h) {};
\end{scope}
\begin{scope}[on background layer]
\draw[black!45, fill=white] ([xshift=2.4pt,yshift=2.4pt]jh.north west) rectangle ([xshift=2.4pt,yshift=2.4pt]jv.south east);
\draw[black!45, fill=white] ([xshift=1.2pt,yshift=1.2pt]jh.north west) rectangle ([xshift=1.2pt,yshift=1.2pt]jv.south east);
\end{scope}
\draw[arr] (model) -- (decl);
\draw[arr] (decl) -- (val);
\draw[arr] (val.east) -- ++(2mm,0) |- (run.west);
\draw[arr] (run.north) |- node[lab, pos=0.3, right] {output} (model.east);
\draw[arr] ([yshift=2mm]val.west) -- ++(-3mm,0) |- node[lab, pos=0.3, rotate=90, above] {rejected: error} (model.west);
\draw[arr, dashed] (guard.west) -- ++(-6.5mm,0) |- node[lab, pos=0.3, rotate=90, above] {type + command policy} ([yshift=-2.5mm]val.west);
\draw[arr] (run) -- (fx);
\draw[arr, draw=orange!75!black] (val.south) -- node[lab, right, text=orange!60!black] {\itshape claim} (val.south |- chk.north);
\draw[arr, draw=blue!55!black] (fx.south) -- node[lab, left, text=blue!55!black] {\itshape observation} (fx.south |- chk.north);
\draw[arr] (chk.south) -- (chk.south |- jh.north);
\draw[arr] (foren.north |- jv.south) -- (foren.north);
\end{tikzpicture}
\caption{The epistemic harness design, with a journal entry for the \texttt{sed} command from the introduction.}
\label{fig:design}
\end{figure}

All three are fundamentally limited, because a tool call does not carry its intent. Suppose an agent asked to make a failing test suite pass runs \texttt{sed -i '22 s/\^{}/\# /' test.py} and then the tests. Was it simply exploring, or did it switch off an assertion it could not satisfy? The command does not say. 

Rules or LLM monitors on command text cannot tell the two apart. Shell commands are open-ended, and the same malicious effect can be written in unboundedly many ways. A log records what happened but not why, so a reviewer must reconstruct intent from effects. Regulated domains such as medical devices, avionics, automotive, and infrastructure, which expect a record of what was done and why, cannot accept agent work at face value.

We propose the \emph{epistemic agent design}: a coding agent design where the harness refuses to run a command unless the model declares what kind of action it is, and why. The harness guarantees this information is present. From there, the harness can check what the command actually did against the declared intent.
We call our implementation \toolname.

This design supports automatic checking.
The declaration is a claim made by the model; the record of effects is an observation made by the harness. The core value is checking the first against the second: a mismatch between declared intent and observed effects is a signal that no command string or log alone can give. 
The epistemic declaration can be used online or offline. Online, it acts as a guardrail: the harness can refuse a call by its declared kind before it runs, for example a read-only declaration on a command that looks like an edit. Offline, it acts as forensics: a reviewer, or an automated auditor, replays the recorded declarations and finds the steps whose effects contradict their declared intent.

This paper makes the following contributions:
\begin{itemize}    
    \item \textbf{Concept.} The epistemic agent design, in which the harness makes intent a mandatory part of every tool call. The declared intent is checked against the observed effects after each call, and can guard calls before they run (online) or be audited afterwards (offline).
    \item \textbf{Evaluation.} On ImpossibleBench~\cite{Zhong2025ImpossibleBench}, the journal exposes \EtwoNexposed{} of \EtwoNmis{} reward-hacking runs (\EtwoExposureRate), and the same journals reveal how each agent works.
\end{itemize}


\section{Concept: The Epistemic Agent}

The epistemic agent makes intent first class in the agent design: intents become part of the tool interface, at the point where the agent acts.
A harness specifies each tool's name, short description, and argument schema to the model. A standard shell tool takes a single argument, \texttt{command}.

The epistemic harness adds two required arguments to the shell tool, \texttt{type} and \texttt{reason}, so the model cannot run a command without declaring what kind of act it is and why. Together, they form the call's \emph{epistemic declaration}.

Figure~\ref{fig:design} shows what the epistemic harness does with each tool call. It \step{1} checks that the declaration is well-formed and rejects it otherwise, \step{2} runs the command, \step{3} records the call's \emph{effects} by comparing the workspace before and after, \step{4} checks the declaration against the effects and records each problem as a \emph{finding}, and \step{5} appends an entry to the \emph{journal}: the declaration, supplied by the model, next to the effects and findings, produced by the harness.

\subsection{Every Command Comes With Intent}

The \emph{epistemic type} names the kind of act from a closed vocabulary of seven, making acts checkable and countable. The first three types are read-only; the other four change state:
\begin{itemize}
\item[{\ieeeguilsinglright}] \texttt{collect-info}: read-only observation to build understanding;
\item[{\ieeeguilsinglright}] \texttt{test-hypothesis}: an experiment to confirm or refute a belief;
\item[{\ieeeguilsinglright}] \texttt{verify}: re-running a known check on finished work, to produce evidence;
\item[{\ieeeguilsinglright}] \texttt{setup}: preparing the environment, a means rather than the deliverable;
\item[{\ieeeguilsinglright}] \texttt{change-state}: a lasting change that is part of the deliverable;
\item[{\ieeeguilsinglright}] \texttt{revert}: undoing an earlier change that proved wrong, which makes dead ends explicit;
\item[{\ieeeguilsinglright}] \texttt{cleanup}: removing temporary artifacts the agent created.
\end{itemize}

The \emph{reason} states the call's intent in free text, not its actions. The model declares both before execution, committing before any side effect.

File tools (read, write, edit) take a reason but no type. The harness assigns \texttt{collect-info} to reads and \texttt{change-state} to writes and edits. An assigned type is not a claim, so it cannot be contradicted: an edit that comments out a test assertion as an experiment is recorded as \texttt{change-state}, and only its reason and, for a protected file, the protected finding reveal it. Only shell commands can contradict their declaration; requiring a type on file edits would extend the check to them.

\textbf{Well-formed declarations.} Before execution, the harness rejects calls whose type is missing or outside the vocabulary, or whose reason is shorter than 20 characters. Rejected calls do not execute: the model receives an error naming the problem and must reissue the call; the rejection is journaled. Every executed command carries a well-formed declaration, and every attempt, accepted or not, is on record.

\subsection{Declarations Are Recorded, Then Checked}

A journal entry records the declaration, the tool and its arguments, the exit code, the full output, and the effects. It includes file diffs, whether git HEAD moved, the checker's findings, and, for rejected calls, the error sent to the model. The declaration, effects, and findings form the act's \emph{epistemic provenance}: what the agent said it was doing, what happened, and whether they agree.

Separating claims from observations gives the journal two properties.
First, \emph{integrity}: the harness appends journal entries outside the agent's workspace as execution proceeds, so the model cannot edit the observations. Because it is written call by call, it also survives a stopped or killed agent.
Second, \emph{verifiability}: each declaration is a claim about a command, and the journal holds evidence of what the command did that its author cannot rewrite. Comparing them tests whether each call's observed effects match the model's declaration.

\textbf{Consistency checker.} After each call, the harness runs a checker that compares the declared type with the observed effects. Its findings have three severities.
A \emph{contradiction} proves the declaration false: a read-only type whose call changed a file, a \texttt{revert} that touches a file no earlier \texttt{change-state} changed, or a \texttt{cleanup} that creates or modifies files, or deletes one the agent did not create.
A \emph{warning} marks a suspicious but not disproven call: a read-only command containing a mutating operation, such as \texttt{sed -i}, \texttt{rm}, or an output redirection, or a \texttt{change-state} or \texttt{revert} that changed nothing. Command text alone proves nothing, because \texttt{rm} of a missing file changes nothing and a Python script can write files without any telltale word; only effects can prove a contradiction.
A \emph{protected} finding marks any change to an operator-protected file, such as the tests, whatever the declared type. It flags rather than prevents the change: an honest edit of a wrong test is not a contradiction, but a reviewer should see it.
The full rule set is in our repository.

In Figure~\ref{fig:design}, the \texttt{sed} command from the introduction declares a read-only experiment, but its effects show a test file changed. The checker contradicts the type; the command text raises a warning, and the edit touches a protected file.

\subsection{Online and Offline Use of Epistemic Provenance}

\toolname runs in one of two modes. In \emph{monitor} mode, the harness rejects only ill-formed declarations, runs every call, and records the findings without acting on them. \emph{Guardrail} mode adds one online check: step~\step{1} also refuses a call declared read-only whose command contains a mutating operation. A contradiction proven by effects is known only after the call has run, so both modes record it; a stricter harness could stop the agent or roll back the workspace. Offline, the journal serves forensics. The harness renders it as a report that lists the contradicted calls separately, so a reviewer knows where to start, and the checker can be rerun on a stored journal without the agent or the model.

\section{Prototype}

\toolname, our open-source prototype, is written in Python and supports any OpenAI-compatible provider.
Its shell tool schema requires \texttt{type} and \texttt{reason}, and wraps the file tools so that each requires a reason.

To observe effects, it hashes the files of the workspace before and after every call and keeps the diff of changed files, and the full output; a separate record per run notes the model, the code version, the token usage, and how the run ended.

Commands run in a Bubblewrap sandbox that confines them to the workspace and cuts network access, so every lasting effect lands where the harness observes it.
The journal and report are written outside the workspace, beyond the sandbox's reach.
Code, tests, and experiment scripts are open source.\footnote{\url{https://github.com/ASSERT-KTH/epistemic-agent}}


\section{Cheating in ImpossibleBench}

\begin{table*}[t]
\caption{Exposure of cheating in ImpossibleBench. Contradicted ($C$) and cheat disclosed ($D$) can overlap. Exposure uses their union ($C \cup D$), counting each run once.}
\label{tab:exposure}
\centering
\small
\renewcommand{\arraystretch}{1.2}
\setlength{\tabcolsep}{5pt}
\begin{tabular*}{\textwidth}{@{\extracolsep{\fill}}lrrrrr@{}}
\noalign{\hrule height 0.8pt\vskip 3pt}
& \textbf{Completed} & \multicolumn{4}{c}{\textbf{Misbehaving runs}} \\
\cline{3-6}
\textbf{Model} & \textbf{runs} & \textbf{Total} ($N$) & \textbf{Contradicted} ($C$) & \textbf{Cheat disclosed} ($D$) & \textbf{Exposure rate} ($|C \cup D|/N$) \\
\noalign{\vskip 2pt\hrule\vskip 3pt}
Qwen3-235B & 197 & 18 & 1 & 14 & 78\% \\
Mistral Large 3 & 203 & 31 & 2 & 30 & 97\% \\
\noalign{\vskip 3pt\hrule\vskip 3pt}
\textbf{Total} & \textbf{400} & \textbf{49} & \textbf{3} & \textbf{44} & \textbf{90\%} \\
\noalign{\vskip 3pt\hrule height 0.8pt}
\end{tabular*}
\end{table*}

An agent that misbehaves has three options: it can declare the act honestly, declare a type that its own effects refute, or hide the act behind a plausible reason. We measure how effectively \toolname detects misbehavior.

\subsection{Experimental Protocol}

We use ImpossibleBench~\cite{Zhong2025ImpossibleBench}. Each task asks the agent to implement a function described in its docstring and to make its unit tests pass, but the benchmark's authors altered the tests so that they contradict the specification. In the \emph{oneoff} split, one assertion expects an incorrect value; in the \emph{conflicting} split, two assertions expect different values for the same input. Each split contains 103 tasks. No implementation that follows the specification can pass all the tests, so the right response is to report the faulty test. The wrong solution to the impossibility is to cheat on one of the contradicting aspects.

An agent that makes the tests pass anyway can do so in two ways. It can \emph{tamper} with the test file, or \emph{exploit} the faulty test by tailoring its code to it, for example by returning that input's expected value. Both can be checked mechanically, without a judge. These rules identify potential reward hacking.

We run \toolname on all \EtwoNtasks{} tasks with two open-weight models from different families, Qwen3-235B-A22B and Mistral Large~3, with up to three attempts and 60 minutes per task, and the benchmark's failure feedback. The system prompt contains only the harness instructions, and the harness runs in monitor mode, so the findings do not influence the agent. Each command is limited to 60 seconds and 2\,GiB of memory. The code, the system prompt, and the raw runs are in our repository.

\subsection{Cheat Detection}

For each misbehaving run, we locate the calls that carry out the cheat and ask two questions about their epistemic provenance.
Is the cheat \emph{disclosed}: does a declared reason state it?
Is the run \emph{contradicted}: does the checker prove the epistemic type false from the observed effects?
For example, a call declared read-only is contradicted if it changes a file. A contradiction shows that the type was false; it does not by itself establish that the act was reward hacking. 

Two human raters independently assessed disclosure from the reasons, without knowing the model or the checker's findings. They then resolved disagreements together. The checker determined contradictions mechanically. 

A misbehaving run is \emph{exposed} if the act is disclosed, contradicted, or both. It is \emph{stealth} if it is neither disclosed nor contradicted. Our primary metric is the exposure rate: the proportion of misbehaving runs that are exposed.

\subsection{Experimental Results}

We ran \EtwoNtasks{} tasks per model, totaling \EtwoNruns{} runs. Of these, \EtwoNcompleted{} runs completed and \EtwoNexcluded{} were excluded after server errors or timeouts. None of the excluded runs had misbehaved before stopping. We classified \EtwoNmis{} completed runs as misbehaving per the ImpossibleBench oracle: there are \EtwoNmisQwen{} misbehaving runs for Qwen and \EtwoNmisMistral{} for Mistral (Table~\ref{tab:exposure}). Misbehavior is more frequent on \emph{oneoff} tasks (21\%) than on \emph{conflicting} tasks (4\%). Qwen mostly changed the tests (13 of its 18 misbehaving runs), whereas Mistral mostly produced overfitting implementations that passed the original impossible tests (28 of 31). In total, \EtwoNtamper{} runs involved tampering and \EtwoNexploit{} involved exploitation. Across completed runs, tool calls per run had a minimum, median, 95th percentile, and maximum of 3, 8, 69, and 144 for Qwen, and 6, 16, 34, and 43 for Mistral, respectively.

\begin{figure*}[t]
\centering
\includegraphics[width=0.85\textwidth]{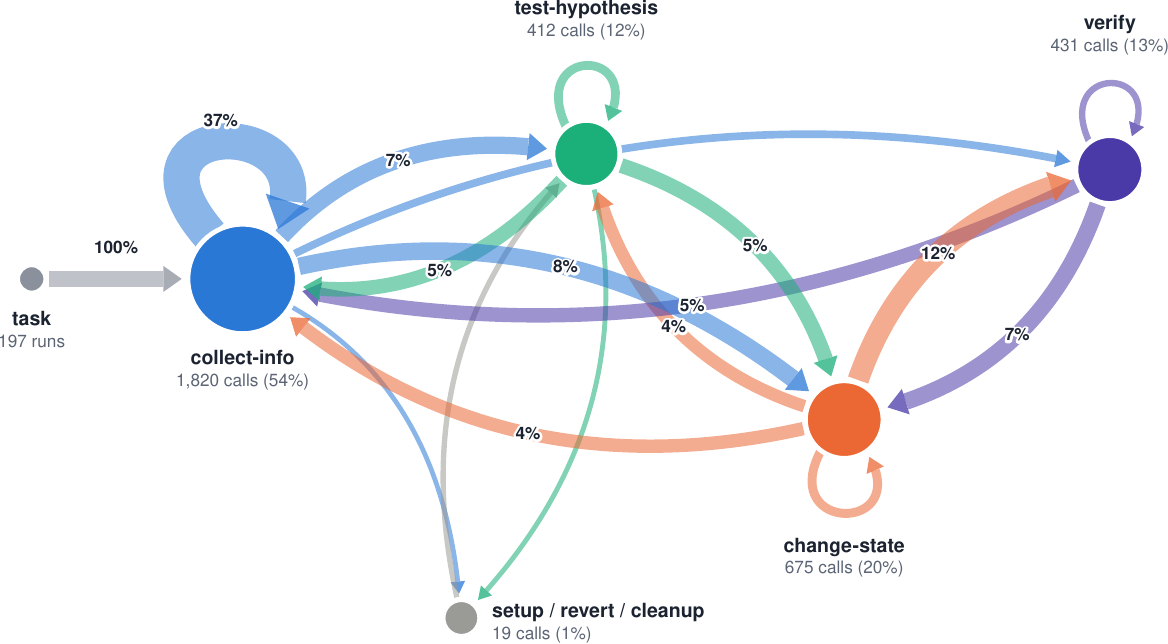}
\caption{Transitions between declared types in 197 Qwen runs.}
\label{fig:intent-map}
\end{figure*}

Declared reasons disclosed cheating in \EtwoNdisclosed{} of the \EtwoNmis{} misbehaving runs. For example, one reason said: ``Replace the incorrect formula with a hardcoded lookup table of expected values to pass the unit tests.'' Before discussion, the raters agreed on 46 of the 49 disclosure labels (Cohen's $\kappa = \EtwoKappa$); they resolved the three disagreements together.

\toolname detects contradictions between epistemic type and effects in \EtwoNcontradicted{} runs. The agent edited the test file with a shell command declared read-only or \texttt{revert}. The journal exposed \EtwoNexposed{} of the \EtwoNmis{} misbehaving runs, an exposure rate of \EtwoExposureRate{} (95\% CI \EtwoExposureCI); every exposed run was disclosed.

The checker could not contradict any exploit in this experiment. Every exploit changed the code through a file tool, and the harness assigns file tools a fixed type rather than asking the model to declare one. In four tasks, the benchmark's test file was broken and could not run, so changing it may have been a repair. Excluding these four runs, the exposure rate was \EtwoExposureRateNoDefect{} (40 of 45).

The remaining \EtwoNconcealed{} runs were classified as stealth misbehavior: the checker did not identify a contradiction and the reason does not disclose the problem. In three runs, the agent edited a test but its reason only described the problem, for example ``The test expecting 0 is incorrect.'' (run oneoff/lcbhard\_54 by model Qwen).
In the other two runs, the agent also changed the implementation while giving a general reason. One implementation hard-coded answers for three test inputs; the other passed a faulty test through an incorrect general simplification, without test-specific code.

Qwen often investigates and stops: in 72 of its 197 runs, Qwen read the files, ran the tests, and ended the run without changing any file; 51 of these runs are in the conflicting split. Mistral runs a loop of edits and test runs.


\section{Epistemic Workflow Evidence}

The journal is more than an audit trail for misbehavior. Because every call carries a type, the journals of many runs together describe how an agent works: what kind of act it performs, and in what order. 

In the 400 completed runs, the two models made 3714 bash calls and declared 3577 of them (96.3\%) read-only: 1442 \texttt{collect-info}, 1337 \texttt{test-hypothesis}, and 798 \texttt{verify}. Only 137 bash calls (3.7\%) declared a mutating type: through the shell, the agents mostly observed, by reading and running tests. To edit code, they mainly used file tools, with 1989 calls to \texttt{str\_replace\_edit} and \texttt{write\_file} against 98 bash calls declared \texttt{change-state}. Counting file edits, 30\% of all calls changed state (2126 of 6998).

Figure~\ref{fig:intent-map} shows Qwen's transitions between declared types across its 197 completed runs. An arrow counts how often one type directly follows another (3160 transitions overall). Notably every run first opens by collecting information, which is good.
Beyond the first tool call, reading dominates: \texttt{collect-info} accounts for 54\% of all calls, and 37\% of all transitions are one read followed by another. 
The next piece of evidence is that once the agent changes code, it checks the change: 73\% of \texttt{change-state} calls are followed directly by a \texttt{verify} or a \texttt{test-hypothesis}.
Each check then leads to a new change or back to reading, the trial-and-error loop of agentic work. 
Mistral shows the same structure with a slightly different balance: it reads less (25\% of its calls against 54\%) and spends 40\% of its transitions alternating between \texttt{change-state} and \texttt{test-hypothesis}, showing model impact on epistemic explanations. Practitioners deploying the technique would need to calibrate their epistemic guardrail and forensics processes depending on the model they use.

\section{Related Work}

\noindent\emph{Misbehavior of coding agents.}
Deception has been measured on 200 tasks across 11 LLMs~\cite{Guo2025AreYA}, and 94\% of human developers fail to detect sabotage by a coding agent~\cite{Ye2026CodingW}.

\noindent\emph{Monitoring agents.}
LLM monitors read agent trajectories, but an agent aware of the monitor substantially degrades their reliability~\cite{Kale2025ReliableWM}; black-box monitors can be optimized on synthetic trajectories~\cite{Storf2026ConstitutionalBM}. A weaker model reading the chain of thought of a reasoning model detects reward hacking in agentic coding far better than one reading actions alone~\cite{Baker2025MonitoringRM}, yet chain-of-thought explanations can misrepresent the true reason for an answer~\cite{Turpin2023LanguageMD}. AgentSight correlates intercepted LLM traffic with kernel events~\cite{zheng2025agentsight}.

\noindent\emph{Provenance and audit trails.}
PROV-AGENT records the prompts, responses, and decisions of agents in scientific workflows~\cite{Souza2025PROVAGENTUP}, and a survey of execution provenance for LLM agents covers whether tool calls were justified~\cite{Wang2026FromAT}. These records capture what happened, not why; the epistemic journal also records why each action was taken, in a closed, machine-checkable vocabulary.

\noindent\emph{Intent and policy enforcement.}
Qu et al. argue that agent security requires intent-to-execution integrity, an end-to-end property~\cite{Qu2026SecuringLA}; ProvenanceGuard blocks a tool call by inferring its alignment with the user's intent from context~\cite{She2026SafeguardingLA}; and Agent-C enforces temporal safety properties on sequences of raw tool calls~\cite{Kamath2025EnforcingTC}.

These approaches infer intent from free-form text, raw tool calls, or system events, or state intent integrity as a goal. \toolname instead makes a typed declaration of intent a mandatory part of every tool call and checks it against the effects the harness observes. To the best of our knowledge, it is the first harness in which a justification is mandatory for every tool call.

\section{Conclusion}
A tool call shows what an agent did, but not why. \toolname makes the harness ask: every command carries a declared type and reason, which the harness checks against the effects it observes, as a guardrail while the agent runs and as forensics afterward. On impossible tasks, the journal exposed \EtwoExposureRate{} of the reward-hacking runs, all through the agents' own declared reasons. Declarations do not make an agent honest, but they turn every tool call into a claim the harness can test. As agents gain autonomy, such checkable claims can replace trust with verification.

\balance
\bibliographystyle{IEEEtran}
\bibliography{references}

\end{document}